# Collective Awareness Platforms and Digital Social Innovation Mediating Consensus Seeking in Problem Situations

Atta Badii[1], Franco Bagnoli[3,5], Balint Balazs[4], Tommaso Castellani[2], Davide D'Orazio[2], Fernando Ferri[2], Patrizia Grifoni[2], Giovanna Pacini[3], Ovidiu Serban[1], Adriana Valente[2]

[1]Intelligent Systems Research Laboratory, University of Reading, UK
[2]IRPPS-CNR, Rome, Italy
[3]Department of Physics and Astronomy and CSDC, University of Florence, Italy
[4]Environmental Social Science Research Group (ESSRG), Hungary
[5]INFN, Sez. Firenze

`giovanna.pacini@unifi.it;franco.bagnoli@unifi.it`

**Abstract.** In this paper we show the results of our studies carried out in the framework of the European Project SciCafe2.0 in the area of Participatory Engagement models. We present a methodological approach built on participative engagements models and holistic framework for problem situation clarification and solution impacts assessment. Several online platforms for social engagement have been analysed to extract the main patterns of participative engagement. We present our own experiments through the SciCafe2.0 Platform and our insights from requirements elicitation.

## 1 Introduction

Collective Awareness Platforms (CAPs) are applications based on Internet or mobile communication, scaffolding on social networking for supporting communities by delivering new services, building innovative knowledge, promoting collective intelligence. The final goal of CAPS is the promotion of more sustainable lifestyles and inducing transformative social innovation [1]. Often such 'voluntary model' is conceptualized as a collaborative commons paradigm as it bypasses the capitalist markets and relies on zero marginal cost [2]. Many applications are devoted to real actions, beyond simple knowledge-sharing, for instance by promoting energy saving (e.g., carpooling, food sharing, buying groups) and essentially harness communication among people [3]. With regard to the optimization and potentials of ICT-enabled spontaneous, massive and collective citizen involvement the concept of crowdsourcing has been recently defined as a process of accumulating the ideas, thoughts or information from many independent participants, with aim to find the best solution for a given challenge [4].

Within the SciCafe2.0 project we are setting up an observatory of crowdsourcing (European Observatory for Crowdsourcing) devoted to participative engagement, in the spirit of the Science Café movement [5,6]. We are studying the information flow in a participative event, and what the psychological and social components beyond the individual participation are. We are also actively experimenting on such phenomena, through specific set-up and developing a platform for supporting participative actions. Finally, we are supporting the Science Café movement, through specific actions and by means of experiments on mixed real-life and Internet-based participative models. Based on this, we present our approach to requirements elicitation that proved to be helpful co-creating robust, scalable and sustainable solutions; then we turn to social-cognitive patterns of social collaboration; and finally analyze patterns of online participative engagement and tools that help such collaboration.

## 2   Holistic Framework for Problem Situation Clarification and Solution Impacts Assessment

One of the most challenging obstacles to resolving problem situations is ambiguity in the problem situation that can cloud the issues and make it difficult to identify the causal roots of the problem. Thus the first priority should be to find a way to establish shared sense making about the problem by overcoming the likely barriers such as the conflicting motives of each (sub) group, their sub-languages, metaphors, subjectivity and any myths and causal fantasies. This process of consensus solution seeking requires dialogue and methodologically-guided elicitation and analysis of the values and priorities of implicated stakeholders.

To empower the stakeholders to achieve a more objective insight about the interplay of influences in the problem space we are reminded that things are most likely to remembered and defended as personal interests worth protecting only in the contexts that they are deemed significant by human beings according to their personal and/or social constructs. Accordingly, to work towards a solution, the contexts of the most valued interests of each implicated sub-group have to be made explicit so as to identify both distinct and shared values and possible trade-offs in specific (sub)contexts. This will pave the way for areas of (inter)subjectivity and (dis)agreement to be delimited within specific (sub)-contexts so as to facilitate consensus solution building.

UI-REF which stands for User-Intimate Integrative Requirements Elicitation and Usability Evaluation Framework [4] is a normative ethno-methodological framework to support problem situation disambiguation, requirements prioritisation and user–solution usability relationship evaluation. As such UI-REF incorporates other methods and instruments, such as empirical ethnographic approaches, cultural probes, laddering, online self-report, action research, nested-video-assisted situation walkthrough, virtual user and gaming-assisted role-play approaches to help reduce the ambiguities. This is achieved by identifying, and de-limiting the areas of disagree-ment and conflicts of interest and concluding the contextualised priorities of the stakeholders in the (sub) problem space(s) where consensus solutions can emerge, endure and thrive to pave the way for increasingly more robust scalable and sustainable solutions to be co-creativity established through deeper engagement as mediated by CAPS tools and digital social innovation.

Various methodologies have been proposed for usability requirements and evaluation and impact assessment. The UI-REF methodological framework is outlined here as one of the possible strategies to elicit and prioritise requirements and ensure maximum possible replicability potential for the resulting solution as well as optimal trade-offs to re local /global and immediate/downstream impacts.

As the relationship between the stakeholders, the problem situation and an emerging solution will evolve over time and the solution needs to be re-visited so as to remain dynamically responsive to evolving realities and relationships of the situation, it follows that there is a need for a Dynamic Usability Evaluation and Holistic Impact Assessment Framework, e.g. the Dynamic Usability Relationship-Based Evaluation (DURE) method [7,8] which takes account of the dynamic relationship that can develop between the stakeholders and the solution.

## 3     Participative Engagement Models

People collaborate for several evolutionary human biology reasons beyond the acting for themselves. Firstly, the genetic component of collaboration implies that collaborating with others is beneficial: even if it is costly or detrimental for the collaborator. This is the main reason for the collaboration in social insects (and in some other animal) and for kin caring - an effective strategy in a small Neolithic village that may lead to quite surprising effects on a highly connected society like ours. Secondly, sexual selection drives the appearance of ornaments (like the peacock tail), which seems useless or even deleterious for survival, but are fundamental for finding a mate. Thirdly, the origin of our intellectual capacities is based on alliances and power. The way this goal is implemented is through rather sophisticated mechanisms of understanding other's wishes (the theory of mind), which is lacking in social-impaired individuals (notably, autistic or suffering from Asperger syndrome). Fourthly, group selection and natural forms of loyalty to our in-group (accompanied with fierce hatred against out-groups) is limited by our cognitive capacities. We apply different heuristics when facing a chat group (4-5 people) or a small group (up to 10-12 people) or a crowd. All in all, we do not generally act following a deep reasoning, but rather applying "rules of thumb" (heuristics) that were successful in our recent (evolutionary speaking) past.

How these heuristics determine our behavior in the Internet world, the propensity towards collaboration, the importance we assign to privacy and reputation, are among the main subjects of our investigations. In particular, we are interested in how they modify when passing from the "real life" world made of physical contacts to the cyber-world, which is missing many of the non-verbal messages we most often rely on. We rely on a tentative classification based on four types and on three functions [7,8]. As for types, we defined four categories as follows:

1. Tools are components used in online participatory activities;
2. Toolkit is a collection of tools that are used in online participatory methodologies;
3. Technique/application is a tool/toolkits put into action (implemented tool/toolkit);
4. Method is a combination of tools, toolkits, techniques put together to address defined goals.

Functions are related to

1. Telling (receive and provide information);
2. Enacting (Discuss, Deliberate, Propose, Vote);
3. Making (Share projects, Co-design projects, Collective problem solving, Share goods).

Such a classification allows to group on-line participatory platforms basing on their primary functionality, identifying 10 paradigms or building blocks of on-line participation [9]: INIP – Interactive Information Provider; AST – Ask-Tell; CODI – Collective Discussion; DIREP – Discussing for Reaching Power Nodes; REP – Reaching Power Nodes; COST – Consulting Stakeholders; SHAGO – Sharing Goods; MAP – Mapping; CODE – Co-Design; COPS – Collective Problem-Solving. These paradigms are considered as 'bricks' with which real participatory platforms are composed.

## 4     The SciCafe2.0 Platform

In the SciCafe2.0 project [10] the main goal is to set-up an observatory of crowdsourcing devoted to participative engagement in the spirit of the Science Café movement. In particular, we aim at the promotion of Science café networks through a supporting agency, the extraction of scenarios and best techniques, the use of this or similar methodologies (like the world cafés) beyond science, and the development of a web interface for supporting this type of communication. We are also interested in the cognitive basis of cooperation, participation and the emergence of collective intelligence.

The technological part of the project was devoted to the development of tools for promoting the combination of different kinds of services for online communities. We profited of the PLAKSS (PLAtform for Knowledge and Services Sharing) framework, developed by CNR [11, 12]. Such a platform was devoted to:

- Set and model the community specifying the characteristics of its participants. PLAKSS can model both people and virtual agents (organizations, devices, etc.) as members of the community.
- Support the different types of collaboration that can occur in Web 2.0 integrating external resources like Google and other social networks:

- content-based: people collaborate sharing content.
- group-based: people collaborate gathering around an idea or interest.
- project-based: people work together on a common task or project such as a development project or a book.

- Set inferring rules for acquiring knowledge and for studying interactions between members of the community. The knowledge can support the modelling and management of complex processes.

  The framework includes functionalities:

- Create a new community specifying the profiles and the information to manage for the members of the community.
- Instantiate the community managing the information, documents and data of the members.
- Support different activities of the community members, like hangouts meetings.
- Manage digital libraries.
- Manage exchange of information and interaction between members
- Share and propagate knowledge between members.

The PLAKSS framework has been used for instantiating the SciCafe2.0 platform. The SciCafe2.0 platform is conceived as a participatory crowdsourcing platform that allows people and organizations to be active actors, playing both the roles of problem and solution providers. It acts as a multiplier of knowledge and innovation:

- Aggregating and making it possible to easily access and share services, information and knowledge already available via pre‑existing tools in an organized and unified manner;
- Enabling users to create their personal repository of services, information and knowledge, that can be shared with other users.

Since the main purpose of the SciCafe2.0 Project is to foster communities dialog and inquiry on specific topics, its members usually need to create a collaborative dialogue and to share knowledge and ideas. For this purpose, the Scicafe2.0 platform implements the dialog:

- Managing Science Café events using the Hangout on-line conference.
- Including in the World Café tool the Hangout on-line conference, and its functionalities such as chats.
- Providing and integrating functionalities allowing users to manage discussions organized in different tables, using and sharing posts, documents, images, videos.
- Providing and integrating functionalities for managing a blackboard for collecting and organizing opinions, or forms (integrated by Google) for managing questionnaires and data collections.

For showing the current development of the platform we introduce how the platform implement the virtual World Café meeting. This kind of virtual meeting is structured in discussion tables specified and configured by the organizer of the World Cafe.

More than one table can be defined. A control panel allows to the "table chair" to start (or re-start) the table discussions. When a table starts, a hangout event is open. Users can play different roles in a virtual World Café meeting:

- Organizer of the World Café
- Table chair
- Participant to a table
- Public of the event

Depending on their role, users can have different views on the defined tables and can play different actions. In particular, an organizer manages the different tables; s/he can assign or change assignment of chairs to the tables. A table chair, when a hangout event is open, can start the table inviting participants; s/he can decide the date and time of the table meeting, or can immediately start and follow the table. Each participant to a table can contribute at the discussion directly by voice, by chat, with opinions on post-it, but can also follow discussions of the other tables (changing her/his role, from "participant" to "public" of the event), accessing the different hangouts live transmission by YouTube users.

Each meeting is also recorded in a video registration of the table on YouTube. The table chair can specify the authorized audience for the registration. Different levels of privacy are managed. In fact, a virtual World Café meeting can be public (and followed by all people, connected to YouTube), or the registration can be restricted to a small group as for example the SciCafe2.0 group that represents the authorized audience.

The World Café tool allows users to organize the space of the blackboard in different areas according to the different aspects or objectives that are discussed in the World Café meeting and need to be modelled. All users can write a post-it putting it on the blackboard and when necessary moving it (according to established semantics) in the different areas of the blackboard shared among participants at the meeting. The blackboard and its content can be saved for the inclusion in the activity stream containing the World Café meeting as one of the activities. After 20 minutes, participants move to another table and add to the content on that table's paper. At the end of the discussion in the table, the documentation (blackboard, forms…videos) of the World Café meeting is automatically collected and recorded into the task having the same title of the World café. Each virtual World café meeting is part of an activity in the SciCafe2.0 platform and more than one virtual World café meeting is usually contained in the same activity. The virtual World café meeting allows users to participate to the stream related to world cafés directly from the activities.

The Citizens' Say Knowledge Exchange is used by the SciCafe2.0 Platform to provide the required access to external knowledge and additional functionalities such as recommendations, Keyword Extraction, Named Entity Recognition, Text Enhancement (Annotation) as well as a parametric description of the way citizens have responded to a participative engagement session – as required and envisaged within the scope of the SciCafe2.0 project. Thus the Citizens' Say Knowledge Exchange provides access to external repositories of information (e.g. DBpedia) and also makes recommendations to the SciCafe2.0 users; suggesting activities/events depending on each user's specific interest (relevant profile) and activities description. This allows the SciCafe2.0 tool to search for individuals, organizations or events that are present in the external Knowledge Repository. One other feature of the Citizens' Say Knowledge Exchange is the Annotation tool, which provides enhanced text information or links, by linking important entities to Wikipedia or DBpedia articles.

## 5   Community Engagement and Requirements Elicitation

EU-level policy supports engagement. The need for stakeholder engagement and transparent dialogue with citizens is clearly articulated in Article 11 of the TEU and the White Paper on European governance (2001). Being the primary customer for SciCafe2.0 project DG Connect has developed its own inclusive approach to the involvement of stakeholders into policies, programmers and services. SciCafe2.0 project regards stakeholder engagement as a process that encompasses relationships built around one-way communication, basic consultation, in-depth dialogue and working partnerships. SciCafe2.0 also developed a Stakeholder Outreach Reference Document to guide co-development of a content marketing plan, which helps to improve the engagement processes. In the document we distinguish four main stakeholder groups as amplifiers, brokers and the medium of the message to the majority:

- Real Communities with Real problems: the project develops and deploys the SciCafe2.0 platform which can be tested in solving community difficulties.
- Partner Communities such as for example CAPS projects, Network of Science Cafes, Responsible Research & Innovation communities, European Innovation Partnerships, The Living Knowledge Network of Science Shops.
- Gateway Networks such as for example The European Network of Regions (ERRIN).
- Public Administration and Policy Making: e.g. DG Connect, Various other Public Institutions such as e.g. Local Authorities. Municipalities.

Our Citizens' Say Participatory Engagement Tool Stakeholders' Requirements Workshop took place in Brussels on the 27th of March, 2014 organized by the SciCafe2.0 Consortium and DG CONNECT. The workshop offered open parallel sessions with stakeholder sub‑groups around small tables discussing their requirements facilitated by members of the SciCafe Consortium. The event was a success with an attendance of approximately 20 persons. The participants included representatives from DG CONNECT, European Regions Research and Innovation Network (ERRIN), Vrije Universiteit Brussels, Responsible Research and Innovation Projects, amongst other persons of interest. The workshop managed to specify the features best valued by the various stakeholders for the Participative Engagement Tool.

From the Citizens' Say Participatory Engagement Tool Stakeholders' Requirements Workshop the SciCafe2.0 Consortium managed to specify stakeholders' requirements for the Participative Engagement Tool. The most important requirements specified from this workshop are that the Tool needs to be simple to use, comforting, the user must be able to set different privacy levels and the tool must also be multilingual.

The SciCafe2.0 Consortium ran a session for practitioners and academics involved and interested in participatory engagement activities. It was held as part of a conference on Innovative Civil Society organized in Copenhagen by the Living Knowledge network of science shops over 9th to 11th April 2014. The session was entitled Scientific Citizenship: Deepening and widening participation and raising the quality of debating and decision making. The objective of this session was to specify more requirements and features best valued by the potential adopters for our Participative Engagement Tool. The workshop was based on Metaplan methodology and aimed at eliciting enablers and barriers from the participants to take part in on-line discussions. The workshop generated a wide variety of insights regarding user requirements; the observation we would like to draw attention to was the repeated emphasis on the social dimensions and constraints: synchronicity, emotions, resonance, collaboration, attendance, reputation and reaching consensus.

From the 1st to the 2nd of July 2014 the SciCafe2.0 Consortium attended the CAPS 2014 Conference held in Brussels. During the first day, 1st of July, a session was held by SciCafe2.0 entitled Citizens' Say: Have Your say! This session was split into two spaces. One was used to present the SciCafe2.0 Project and Citizens' Say Platform and the second one was a "hands‑on‑session". Both parts of the session as held at the CAPS2014 Conference were attended by a smaller audience than previous workshops but there was an open plan setting and people just dropped in and out so the total audience was larger than that at any time. Participants could explore and experiment with the platform on their computers in the room, assisted by the SciCafe 2.0 Consortium members who continued to engage with the audience.

## 6  A real case study "Science with and for Society Observatory"

Within SciCafe 2.0 project we have made a comparison of existing on-line participatory methodologies [13], we have implemented and edited a Handbook of Online Participatory Methodologies [14] in which some paradigms of on-line crowd-sourcing participatory methodologies are proposed, based on the analysis of online platforms. The emerging results contributed to the development of the SciCafe 2.0 platform.

The SciCafe 2.0 platform integrating the Citizens' Say Knowledge Exchange Tool provides for stimulating participation/cooperation. Basing on the results of this preliminary work a Delphi-based model of collective participation and knowledge building in the decision making processes was designed and implemented on the SciCafe 2.0 Platform. The aim of the model was to explore the potential of CAPS in participative policy-making, directly connecting with real social contexts and including relevant social actors. We implemented the participatory model to a real case study, in our case the "Science with and for Society Observatory" of the Second Municipality of Rome. The participatory process was designed from the beginning as a combina-tion of online and offline activities and It is based on the collective participation of a multitude of different actors, including policy makers, experts, citizens. As a typical Delphi model, our process is divided in different steps, and combines off-line and on-line activities

This experiment was a success, more of three hundred people have participated at the first plenary meeting of the Observatory, and many of them decided to continue the participatory process participating at the works of the different groups and at the online activities hosted by SciCafe2.0 platform. After the implementation of the participatory process, we performed the validation by means of an online workshop in which we applied the Delphi methodology within the RE-AIM Framework.
The Delphi-based validation workshop involved 10 panelists from different categories (policy makers, researchers, science museums, schools and citizens) who discussed the effectiveness of online participatory decision making as well as the advantages and specifics of the different participative instruments and. we obtained a high level validation of the participatory model; in fact we received 8 recommendations regarding the implementation of a participatory model in order for the participatory model to be successful.

In our experience we can say that Scicafe 2.0 and its participatory model is a suitable for change in the dynamics of social innovation especially those relating to participatory methodologies aimed to citizens involvement and bottom up actions, it facilitates and entices users to the implementation of participatory methodologies applicable to different fields and walks of life, becoming a powerful tool for public engagement within the broader dynamics of social innovation within the macro changes that new media favor, but also stand out in the social dynamics, economic policies, effectively making them a major catalyst for change.

## 7   Lessons learnt and insights arising from the participatory engagement sessions

- Human resources are the real limitations in organizing moderator support for the participatory engagement sessions. In order to furnish a satisfactory interaction, people connected from remote need to receive timely responses from the online moderator. In practice, this means that one needs two moderators, one for the online engagement and one for the face-to-face interactions. It was concluded that the availability of some kind of automatic moderation facility should be explored as an added value, for instance using a portable device (a tablet) for online moderation by a single moderator.
- The inclusion of various instruments within the SciCafe2.0 platform, such as Google Hangouts, requires a particular attention to the third parties' policies (e.g. access, data storage, privacy, etc.). Indeed, one can need to visualize copyrighted material (like for instance pieces of movies), the recording of which will be blocked by Google.
- Non-verbal communication (i.e. emotion) was considered as an essential aspect both in written and oral discussions. Therefore, the implementation of some kind of emotional feedback within the SciCafe2.0 platform would be an added value. One possibility is that of adding emoticons, like in WatsApp and Google chat.
- For the Hangout discussion sessions, some capability for regulating the turn taking in speaking was considered as desirable.
- Enhanced support for the users maintaining an overview of the discussion themes and threads, e.g. by way of some graphic tool was also considered as a helpful feature. We already implemented the threaded discussion and the use of colors for distinguishing recent from old messages. A further possibility would be to add separate "rooms" for the discussion and a kind of "wall" where the main ideas arising from the rooms' discussion can be publicly posted. This was part of the original design of the interface, whose implementation was however delayed for technical problems.

## 8  Conclusions

In this paper the authors have presented an account of their studies in the area of Participatory Engagement models specifically addressing the aspects of Collective Awareness Platforms and Digital Social Innovation Mediating Consensus Seeking in Problem Situations. The paper has explored the various influences at play in societal problem situations including socio-psycho-cognitive, social engagement models, constructs and the situated cultural, and ambiguity challenges of the problem environment as well as the methodologically-guided means of reducing ambiguity, thus reducing and delimiting the contexts where there is disagreement and in doing so increasing agreement including about disagreements - towards consensus solution co-creation. The paper also briefly describes the UI-REF Framework for problem situation disambiguation and Requirements prioritization. The SciCafe platform, including the Citizens' Say tool, is featured as an example of Engagement Platforms and the World Café as an example of a Participatory Engagement Model. The paper concludes with an account of the SciCafe 2.0 user requirements elicitation and community engagements and the resulting insights shared.

**Acknowledgements.** The SciCafe2.0 Consortium wishes to acknowledge the support from the European Community DG Connect under ICT-611299 the Collective Awareness Platforms

## References

1. Sestini, F.: Collective awareness platforms: Engines for sustainability and ethics. Technology and Society Magazine, IEEE, 31(4), 54-62. (2012)
2. Rifkin, J.: The zero marginal cost society: The internet of things, the collaborative commons, and the eclipse of capitalism. Macmillan, (2014).
3. Bagnoli, F., Guazzini, A., Pacini, G., Stavrakakis, I., Kokolaki, E., & Theodorakopoulos, G.: Cognitive structure of collective awareness platforms. In Self-Adaptive and Self-

Organizing Systems Workshops (SASOW), 2014 IEEE Eighth International Conference on (pp. 96-101). IEEE (2014)
4. Guazzini, A., Vilone, D., Donati, C., Nardi A., Levnajić Z. Modelling crowdsourcing as collective problem solving. Scientific Reports 5, Article number: 16557 (2015), doi:10.1038/srep16557. URL: http://www.nature.com/articles/srep16557
5. Dallas, D.: The Cafè Scientifique. Nature, 399, 120.doi:10.1038/20118, (1999)
6. Pacini G., Bagnoli F., Belmonte C., Castellani T.: Science is ready, serve it! Dissemination of Science through Science Cafè, in Quality, Honesty and Beauty in Science and Technology Communication, PCST 2012 book of papers, edited by M. Bucchi and B. Trench, Observa Science in Society, (2012)
7. Badii A.: User-Intimate Requirements Hierarchy Resolution Framework (UI-REF): Methodology for Capturing Ambient Assisted Living Needs. Proceedings of the Research Workshop, Int. Ambient Intelligence Systems Conference (AmI'08), Nuremberg, Germany, November 2008
8. Badii A.: Online Point-of-Click Web Usability Mining with PopEval-MB, WebEvalAB and the C-Assure Methodology. Proceedings of Americas Conference on Information Systems (AMCIS 2000), University of California, Long Beach, (2000).
9. Sanders E.B.-N., Brandt E., Binder T.: A framework for organizing the tools and techniques of participatory design. Proceedings of the 11th biennial participatory design conference. ACM, pp 195–198, (2010)
10. http://scicafe2-0.european-observatory-for-crowdsourcing.eu/
11. Ferri, F., Grifoni, P., Caschera, M.C., D'Ulizia, A.: Praticò, C. KRC: KnowInG crowdsourcing platform supporting creativity and innovation, AISS: Advances in Information Sciences and Service Sciences, Vol. 5, No. 16, pp. 1 -15. (2013)
12. Grifoni P., Ferri F., D'Andrea A., Guzzo T., Pratico C. , SoN-KInG: A digital eco-system for innovation in professional and business domains Journal of Systems and Information Technology, 16 (1) , pp. 77-92. (2014)
13. Castellani T., D'Orazio D., Valente A.: Case studies of on-line participatory platforms, IRPPS Working Papers 67, (2014)
14. Castellani T., Valente A., D'Orazio D., Ferri, F., Grifoni P., D'Andrea A., Guzzo T., Balazs B., Bagnoli F., Pacini G., Badii A.: Handbook of on-line participatory methodologies, SciCafe 2.0 Deliverable D4.1, (2014)